\documentclass[aps,prd,twocolumn,10pt]{revtex4}
\usepackage{xcolor}
\definecolor{Gray}{gray}{0.85}
\definecolor{LightCyan}{rgb}{0.88,1,1}
\usepackage[colorlinks,linkcolor={red},citecolor={blue}]{hyperref}
\usepackage{eurosym}
\usepackage{graphicx}
\usepackage{amsmath}
\usepackage{amssymb}
\usepackage{amsfonts}
\usepackage{bm}
\usepackage{hhline}

\usepackage{multirow}

\def\be{\begin{equation}}
\def\ee{\end{equation}}
\def\bea{\begin{eqnarray}}
\def\eea{\end{eqnarray}}

\newcommand{\f}[2]{\frac{#1}{#2}}

\begin{document}
\title{Cosmological implications of Rastall-$f(R)$ theory}
\author{Shahab Shahidi}
\affiliation{School of Physics, Damghan University, Damghan 41167-36716, Iran.}
\date{\today}

\begin{abstract}
Cosmological implications of a generalized Rastall theory where the non-conservation equation is promoted to accepts arbitrary functions of the Ricci scalar is considered. We have obtain the model parameters of the power-law ansatz $f\propto R^\eta$, using $H_0$ and $f\sigma_8$ data sets and show that the generalized Rastall theory could satisfy observational data. The dynamical analysis of the model shows that for $\eta\in(0,1)$, the dust dominated fixed point is lost and as a result the parameter range of $\eta$ should be restricted to $\eta\geq1$. We will also show that the genralized Rastall theory will predict more accelerating universe with larger radius, compared to the $\Lambda$CDM model.
\end{abstract}

%\pacs{}
\maketitle
\section{Introduction}
Einstein theory of general relativity has long been accepted as a main theory describing gravitational interactions. As an application to cosmology, the assumption that the geometry of the universe could be well described by the Friedmann-Robertson-Walker metric is now supported strongly by observations \cite{frwobs} which indicated that the universe undergoes an accelerated expansion. Noting that the gravitational force in general relativity is attracting, a new form of ``dark energy" with negative pressure and repulsive nature would be needed to explain late time observational data. The simplest idea of dark energy comes from Einstein himself, which is adding a cosmological constant to the original theory. Together with the cold dark matter component, accounting for local observations, the $\Lambda$CDM model is now widely accepted as a general ground of cosmological theories \cite{concor}. 

Apart from its successes, the $\Lambda$CDM model suffers from some phenomenological/theoretical problems, including the cosmological constant problem and also the $H_0/\sigma_8^0$ tensions \cite{cosmologicalconstantproblem,hubbletension}. It is then important to investigate other possibilities which we will refer to as modified theories of gravity \cite{modifiedgravityreview}. The modified gravity idea has in fact many branches. One proposal deals with the possibility that accelerated expansion of the universe could be produced via additional degrees of freedom, such as scalar \cite{scalartheories}, vector \cite{vectortheories} or higher spin fields \cite{higherspintheories}. Other possibilities would be to change the gravitational interaction itself, like in massive gravity theories \cite{massivegravity} or using non-Riemannian geometries, like Weyl/Cartan theories \cite{weylcartan}.

One of the widely investigated modifications to the Einstein's general relativity is to promote the Einstein-Hilbert Lagrangian to an arbitrary function of the Ricci scalar which is well-known as $f(R)$ theories of gravity \cite{fRreview}. Several aspects of this theory is investigated in the literature including cosmological implications and black hole solutions. The idea of $f(R)$ theory can also be generalized to include other curvature terms like the Gauss-Bonnet invariant in $f(R,G)$ theories \cite{fGreview} and also matter fields. The last possibility has attracted more attention in recent years, resulting in modification of general relativity like $f(R,T)$ \cite{fRT}, $f(R,Lm)$, \cite{fRLm}, $f(R,T,R_{\mu\nu}T^{\mu\nu})$, \cite{fRTRT}, derivative matter couplings \cite{derivative} and also energy-momentum squared type theories \cite{EMSgravity}. 

The Einstein general relativity theory is written in such a way that the matter sector remains conserved. In 1972, Rastall has proposed a theory in which the conservation of the energy-momentum tensor was relaxed \cite{rastalloriginal} and replaced by the relation
\begin{align}
\nabla_\mu T^\mu_{~\nu}=a_\nu,
\end{align}
where the vector $a_\nu$ vanishes in flat space-times. Einstein assumed minimal substitution as a way to promote special relativity to curved space-times. The theory proposed by Rastall has in fact relaxed the proposal of minimal substitution for writing the conservation equation of matter fields in curved space-time. As a result, every vector field which is constructed from the curvature tensor could do the job. In the original Rastall theory, it is assumed $a_\nu=\lambda\nabla_\mu R$. Rastall theory has attracted many attentions in recent years \cite{attractedtorastall}, specially in cosmology \cite{rastallcosmology}. 
For example in \cite{04074} the authors considered the observational constraint on the Rastall theory and claimed that there is no way that both $H_0$ and $\sigma_8^0$ tensions become better. In fact, making the Hubble tension better will worsen the $\sigma_8^0$ tension and vice versa.

One can also assume more general form of the vector $a_\nu=\nabla^\mu A_{\mu\nu}$, where $A_{\mu\nu}$ is a symmetric second rank tensor with the property that the tensor and its derivatives vanish on flat space-time. For example in \cite{10075}, the authors obtained charged black hole solutions in the case that $A_{\mu\nu}$ contains all terms constructed by the Riemann tensor and its derivatives that produces up to 4 space-time derivatives. In \cite{10100} the authors obtained neutral black holes for $A_{\mu\nu}=fg_{\mu\nu}$ with arbitrary function of $f$. Also, one can assume a Brans-Dicke-Rastall type theories \cite{BransDickeRastall}. In \cite{03229} the authors investigated the cosmic acceleration for a specific function $f=f(R,T)$.

One of the important debates on the Rastall idea is to construct a fully covariant action which leads to the Rastall equations of motion. For example a relationship between Rastall theory and $f(R,L_m)$ theory is considered in \cite{10503} where the authors constraint the Rastall parameter to lie in the range $\lambda\leq0$ and $\lambda\geq1$ using energy conditions. Also, the relation between Rastall theory and k-essence model is considered in \cite{06662} where the authors claimed that both theories produce same solutions. The most successful of these attempts is done with the $f(R,T)$ theory in which the Lagrangian $f(R,T)=R+\alpha T$ is considered as a Rastall Lagrangian \cite{08203andhadi}. It should be noted that there were some other debates which claimed that the Rastall theory is in fact equivalent to the standard Einstein's theory \cite{11500}. In \cite{09307} the authors tried to answer the debate. However, the equivalence of Rastall and GR is a very controversial problem and is one of these days research lines.

In this paper, we consider a generalization of the Rastall theory in which the tensor $A_{\mu\nu}$ can be written as a general function of the Ricci scalar, i.e. $A_{\mu\nu}=f(R)g_{\mu\nu}$. The energy-momentum conservation equation in this case is modified as
\begin{align}\label{eqcons}
\nabla^\mu T_{\mu\nu}=2\kappa^2\nabla_\nu f(R).
\end{align}
One can see that the modified Einstein equation would be
\begin{align}\label{eqeins}
G_{\mu\nu}+\big(\Lambda+f(R)\big)g_{\mu\nu}=\f{1}{2\kappa^2}T_{\mu\nu},
\end{align}
where  $\Lambda$ is the cosmological constant.
It should be noted that in the case $f(R)=\lambda R$, we recover the standard Rastall theory.  As was mentioned before, the same generalization is used to obtain the black hole solutions in Rastall theory. In this paper, we will examine cosmological implications of the model and try to obtain the best fit values of the model parameters using observational data on the Hubble parameter and also the $f\sigma_8$ function. We will restrict ourselves to a specific function of the  form $f=\xi R^\eta$, for better comparison with observations. We will show that for positive values of the parameter $\eta$, the value of $\xi$ up to $2\sigma$ confidence level should also be positive. We will also show that the values of $0<\eta<1$ is not fully appropriate cosmologically, since the matter dominated era could not be occurred in this case. 

The paper is organized as follows. In the next section, we will consider the background cosmology of the theory and show that the theory can satisfy observational data at late times. The universe is however, larger in size compared to the $\Lambda$CDM model due to the presence of the non-conservative term in the theory. In section \ref{dynamicalsystem}, we will consider the dynamical system analysis of the model and prove that the theory does not have a dust fixed point for $0<\eta<1$. In section \ref{cosmopert}, we will analyze matter perturbations of the theory and find the evolution equation of the matter density contrast. We will then use the dynamical equations to find the best fit values of the model parameters using the observational data on the Hubble parameter and $f\sigma_8$ functions. In the last section we will conclude the paper.

\section{Cosmology}
Now, let us consider a homogeneous and isotropic spacetime with the conformal FRW line element of the form
\begin{align}
ds^2=a(t)^2(-dt^2+d\vec{x}^2),
\end{align}
where $a(t)$ is the scale factor and $t$ represents the conformal time. We also assume that the universe is filled with a perfect fluid with energy-momentum of the form
\begin{align}
T^\mu_{~\nu}=\textmd{diag}(-\rho,p,p,p),
\end{align}
where $\rho$, $p$ are the energy density and thermodynamic pressure respectively.

The Friedmann and Raychaudhuri equations can be obtained from \eqref{eqeins} as
\begin{align}
3H^2&+a^2(\Lambda+f)=\f{1}{2\kappa^2}a^2\rho,\\
H^2&+2\dot{H}-a^2(\Lambda+f)=-\f{1}{2\kappa^2}a^2\,p.
\end{align}
Here, $H=\dot{a}/a$ is the Hubble parameter. Also $f=f(R)$ where $R$ is the Ricci scalar which is equal to $$R=\f{6}{a^2}(\dot{H}+H^2),$$ in flat FRW universe. The conservation equation \eqref{eqcons} can also be written as
\begin{align}
\dot{\rho}+3H(\rho+p)=\f{12\kappa^2}{a^2}(2H^3-\ddot{H})f_R,
\end{align}
where $f_R=d f/d R$.

Now let us assume a special form $f(R)=\epsilon R^\eta$ for the function $f$, where $\eta$ is a dimensionless constant and $\epsilon$ is an arbitrary constant with mass dimension $2(1-\eta)$. In this case, the Friedmann, Raychaudhuri and conservation equations reduces to
\begin{align}
3H^2&-\epsilon a^2\left(\f{6}{a^2}(H^2+\dot{H})\right)^\eta=\f{1}{2\kappa^2}a^2(\rho+2\kappa^2\Lambda),\\
H^2&+2\dot{H}-\epsilon a^2\left(\f{6}{a^2}(H^2+\dot{H})\right)^\eta=-\f{1}{2\kappa^2}a^2(p-2\kappa^2\Lambda)
\end{align}
and
\begin{align}\label{consfrw}
\dot\rho+3H(\rho+p)=\f{12\epsilon\kappa^2\eta}{a^2}(2H^3-\ddot{H})\left(\f{6}{a^2}(H^2+\dot{H})\right)^{\eta-1}.
\end{align}
It is worth mentioning that the above set of field equations reduce to the standard Einstein equation with cosmological constant in the case $\epsilon=0$.

Let us assume that the matter sector of the universe consists of dust with equation of state $p=0$ and radiation with $p=\rho/3$, so that
\begin{align}
\rho=\rho_m+\rho_r,\quad p=\f13\rho_r.
\end{align}
Considering equation \eqref{consfrw}, one can see that the combination of dust and radiation are not conserved in general. However, as we know from the late time observations, the radiation abundance of the universe is significantly lower than dust and one could expect that the radiation part of the matter content does not contribute much to the non-conservation equation. As a result, we assume in this paper that the radiation part of the universe is conserved and all the non-conservation factors in \eqref{consfrw} is carried by dust. 
\begin{figure*}
	\includegraphics[scale=0.55]{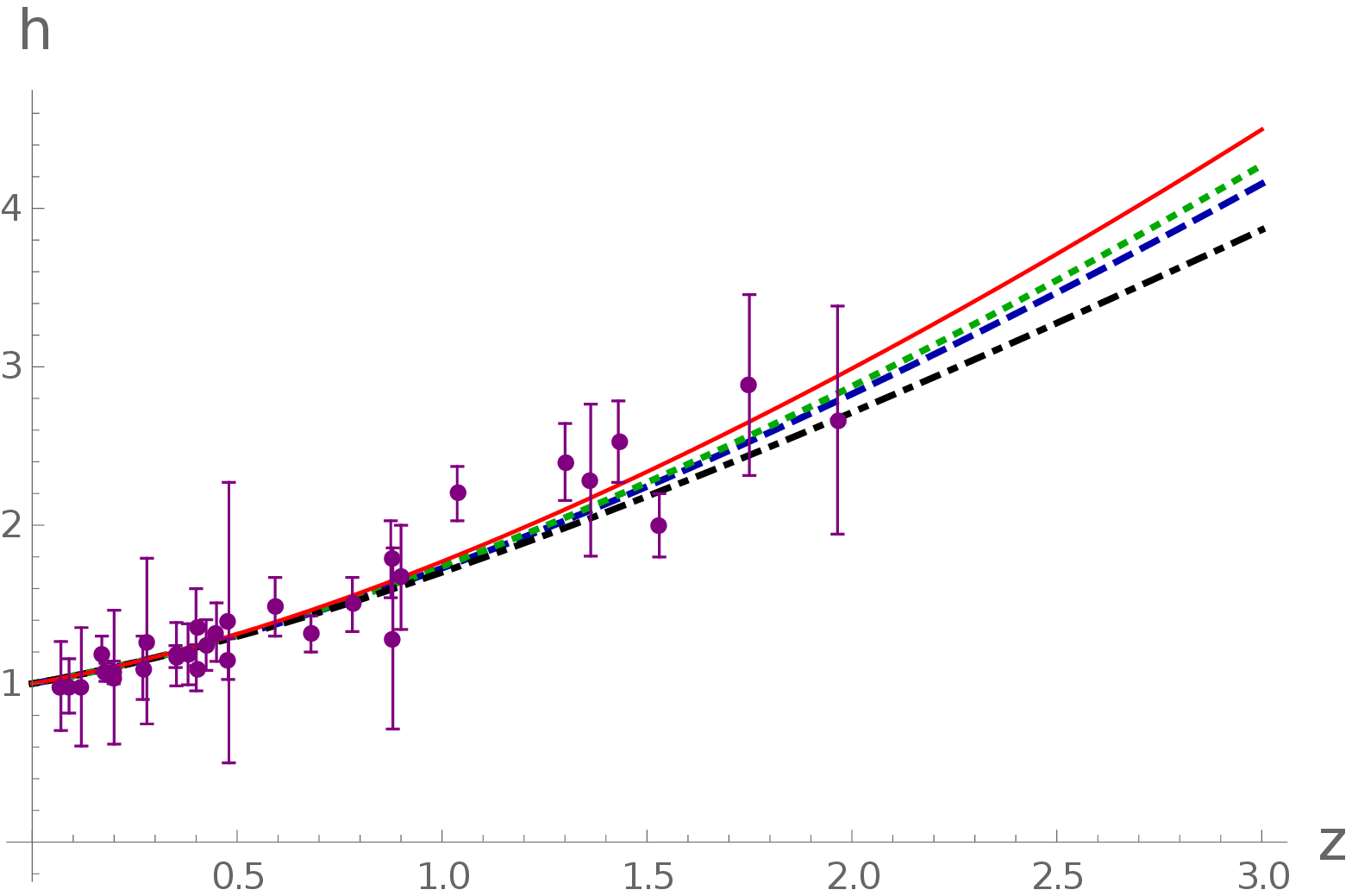}\quad\includegraphics[scale=0.55]{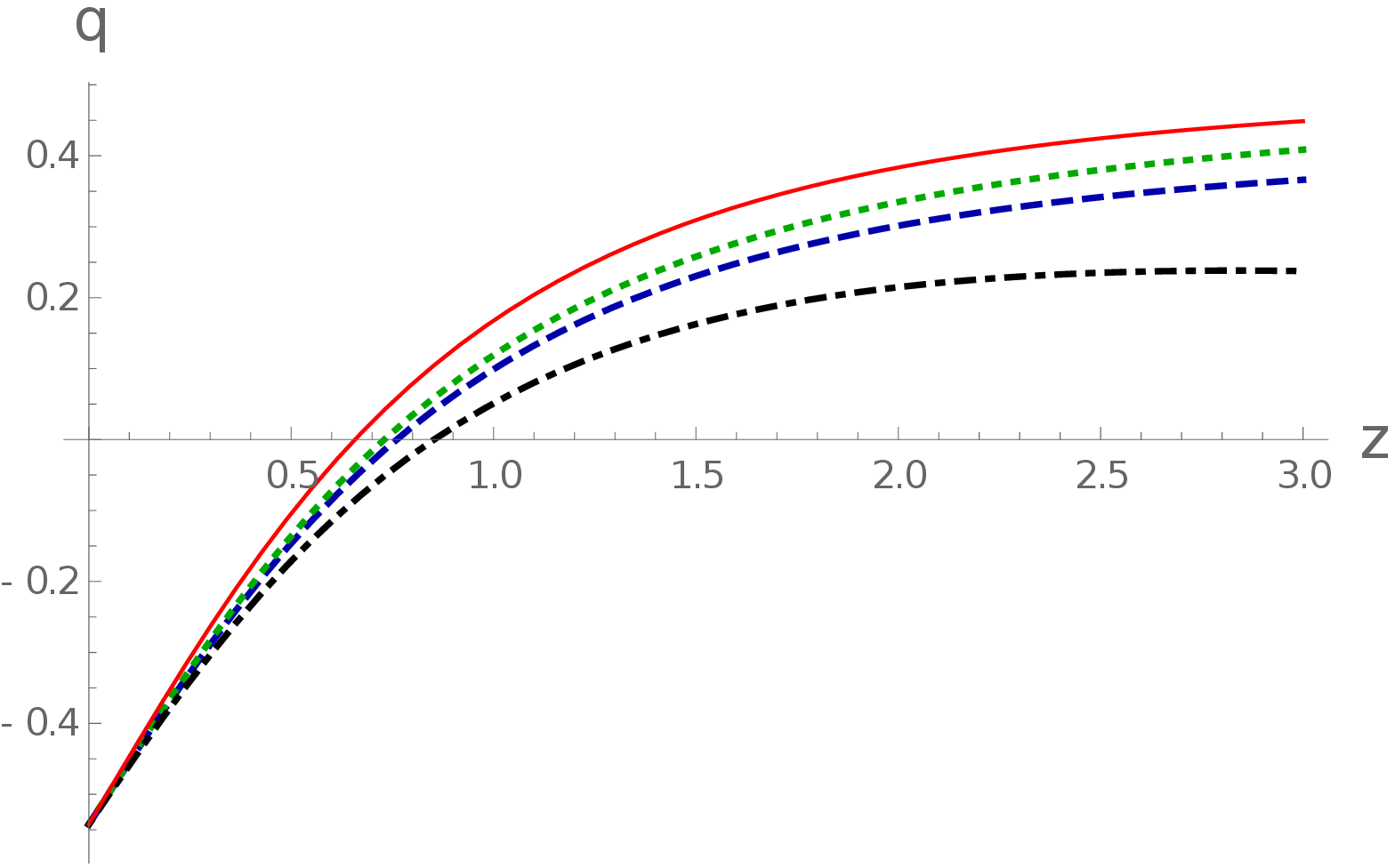}
	\caption{\label{fhubq}The evolution of the Hubble parameter $h$ (left) and the deceleration parameter $q$ (right) as a function of redshift for different values of $\eta=0.6$ (dotted), $1$(dashed), $1.3$ (dot-dashed). The solid red line corresponds to the $\Lambda$CDM theory.  The error bars indicate the observational values \cite{obshubble}.}
\end{figure*}
Defining the following dimensionless variables as
\begin{align}
\tau&=H_0t,\quad H=H_0h,\quad \xi=6^\eta\epsilon H_0^{2(\eta-1)},\\
\bar{\rho}_i&=\f{\rho_i}{6\kappa^2 H_0^2},\qquad \Omega_{\Lambda}=\f{\Lambda}{3H_0^2},
\end{align}
where $i=m,r$ corresponding to dust and radiation respectively, and $H_0$ is the current value of the Hubble parameter, one can obtain the dimensionless Friedmann, Raychaudhuri and conservation equations as
\begin{align}
h^2=a^2\left[\bar{\rho}_m+\bar{\rho}_r+\Omega_{\Lambda0}+\f13\xi\left(\f{h^2+h^\prime}{a^2}\right)^\eta\right],\label{frid}\\
h^2+2h^\prime=a^2\left[-\bar{\rho}_r+3\Omega_{\Lambda0}+\xi\left(\f{h^2+h^\prime}{a^2}\right)^\eta\right],
\end{align}
\begin{align}
	\bar{\rho}_r^\prime&+4h\bar{\rho}_r=0,\label{radcons}\\
	\bar{\rho}_m^\prime&+3h\bar{\rho}_m=\f{\eta\xi}{3a^2}(2h^3-h^{\prime\prime})\left(\f{h^2+h^\prime}{a^2}\right)^{\eta-1}
\end{align}
where prime denotes derivative with respect to the dimensionless time $\tau$.

For the evolution of the radiation density, one can see from \eqref{radcons} that
\begin{align}\label{rad1}
\bar{\rho}_r=\f{\Omega_{r0}}{a^4},
\end{align}
where $\Omega_{r0}=5.3\times10^{-5}$ is the current abundance of the radiation \cite{plank}.

In order to compare the theory with observational data, it is more convenient to work with redshift coordinates, defined as
\begin{align}
1+z=\f{1}{a}.
\end{align}
One can then rewrite the Friedman and Raychaudhuri equations in the redshift coordinates as
\begin{align}\label{eq1}
h^2=(1+z)^2\Omega_{r0}&+(1+z)^{-2}(\Omega_m+\Omega_\Lambda)\nonumber\\&+\f13\xi(1+z)^{2(\eta-1)}h_1^\eta
\end{align}
and
\begin{align}
h^2&-2(1+z)hh^\prime=-(1+z)^2\Omega_{r0}\nonumber\\&+3(1+z)^{-2}\Omega_\Lambda+\xi(1+z)^{2(\eta-1)}h_1^\eta,
\end{align}
where we have defined
\begin{align}
h_1=h^2-(1+z)hh^\prime.
\end{align}
Also, we have used the conservation equation of radiation \eqref{rad1}. We should note that prime denotes derivative with respect to the redshift coordinate, wherever we work on redshift coordinates. The (non-)conservation equation of the dust fluid can be written in the redshift coordinates as
\begin{align}\label{eq2}
(1+z)\Omega_m^\prime-3\Omega_m=-\f13\xi\eta(1+z)^{2\eta}(2h_1+(1+z)h_1^\prime).
\end{align}
It should be mentioned that the value of the cosmological constant abundance $\Omega_\Lambda$ could be obtained from other constant parameters by noting that $h(z=0)=1$. As a result one can obtain from the Friedman and Raychaudhuri equations that
\begin{align}
\Omega_\Lambda=1-\Omega_{m0}-\Omega_{r0}-\f13\xi\left[2-\f32\Omega_{m0}-2\Omega_{r0}\right]^\eta,
\end{align}
where $\Omega_{m0}=0.305$ is the current value of the dust fluid abundance \cite{plank}.

In section \ref{cosmopert}, we will obtain the best estimation of the parameter $\xi$ and also the current value of the Hubble parameter, using observational data on $H$ and also $f\sigma_8$. In table \eqref{tab2}, we have summarized the best fit values together with its $1\sigma$ and $2\sigma$ confidence intervals of the parameters $h_0\equiv H_0/70$ and $\xi$ for three different values of the parameter $\eta=0.6,\,1,\,1.3$. In figure \eqref{fhubq}, we have plotted the evolution of the Hubble parameter and also the deceleration parameter $q$ defined as
\begin{align}
q=(1+z)\frac{h^\prime}{h},
\end{align}
as a function of the redshift for $\eta=0.6,\,1,\,1.3$. The red solid line indicates the $\Lambda$CDM result. The case $\eta=1$ is the original Rastall theory. To plot the figures, we have used the best fit values of table \eqref{tab2}. One can see from the figures that all cases can satisfy the observational data on the Hubble parameter. For redshifts smaller than $z\approx1$, the generalized Rastall theory is identical to the $\Lambda$CDM theory. For larger redshifts, the generalized Rastall theory predicts smalled Hubble parameter which results in larger size of the universe compare to the $\Lambda$CDM model. It can be seen from the figure that as $\eta$ increases, the Hubble parameter decreases, leading to the larger universe for larger $\eta$ values. Also, the deceleration parameter would be smaller than the $\Lambda$CDm model for $z>0.2$. This implies that the generalized Rastall theory predicts more accelerating universe compare to the $\Lambda$CDM model at these redshifts. This is however compatible with the Hubble diagram of the theory. 
\begin{figure}
	\includegraphics[scale=0.54]{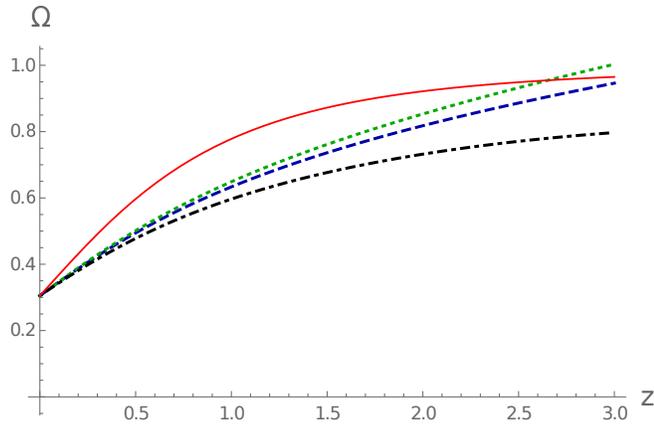}
	\caption{\label{fomega}The evolution of the dust abundance $\Omega_m$ as a function of redshift for different values of $\eta=0.6$ (dotted), $1$(dashed), $1.3$ (dot-dashed). The solid red line corresponds to the conservative $\Lambda$CDM theory.}
\end{figure}

The evolution of the matter density abundance $\Omega_m=\bar{\rho}_ma^2/h^2$ is depicted in figure \eqref{fomega} for different values of the parameter $\eta=0.6,\,1,\,1.3$. It should be noted that the dust is not conserved in the generalized Rastall theory. The red solid curve in the figure corresponds to the conservative $\Lambda$CDM theory where the dust density abundance behaves like $\bar{\rho}_m\propto(1+z)^3$. One can see from the figure that the dust density abundance is lower than the $\Lambda$CDM case. This can be explained by the fact that the matter field converted to geometry due to the non-conservative nature of the matter field. For large enough values of the resdhift, the matter density abundance of the generalized Rastall theory could become higher than the $\Lambda$CDM value. This in fact compatible with the evolution of the deceleration parameter which we have depicted above. From equation \eqref{eq2}, one can see that the right hand side of the equation is always negative. This means that larger values of $\Omega_m^\prime$ corresponds to smaller values of matter converted to the curvature which means lower attraction and more acceleration. 

In the next section, we will consider the dynamical system analysis of the model for better understanding of the evolution of the universe in the generalized Rastall theory.
\section{Dynamical system analysis}\label{dynamicalsystem}
Defining the following matter density abundance
\begin{align}
\Omega_i=\f{\bar\rho_ia^2}{h^2},\quad\Omega_\Lambda=\f{\Omega_{\Lambda0}a^2}{h^2},\quad \Omega_h=\f{h^2}{a^2},
\end{align}
where $i=r,m$, one can write the Friedman equation \eqref{frid} as
\begin{align}\label{dyna1}
1=\Omega_m+\Omega_r+\Omega_\Lambda+\f13\xi\Omega_h^{\eta-1}\left[\f13(4-4\Omega_m-4\Omega_r)\right]^\eta.
\end{align}
The dynamical variable $\Omega_h$ could be considered as a density abundance of the curvature which is the characteristic of the generalized Rastall theory.

By obtaining $\Omega_\Lambda$ from equation \eqref{dyna1}, one can deduce that we have three independent dynamical variables $\Omega_m$, $\Omega_r$ and $\Omega_h$ with dynamical equations
\begin{align}\label{dy1}
\Omega^\prime_m&=\Omega_m\Bigg[3\Omega_m+4\Omega_r-4\nonumber\\&+\f{1}{1+3\xi\eta\Omega_h^{\eta-1}\left(3(4-3\Omega_m-4\Omega_r)\right)^{\eta-1}}\Bigg],
\end{align}
\begin{align}\label{dy2}
\Omega^\prime_r=\Omega_r(3\Omega_m+4\Omega_r-4),
\end{align}
\begin{align}\label{dy3}
\Omega^\prime_h=\Omega_h(3\Omega_m+4\Omega_r),
\end{align}
where here, prime denotes derivative with respect to $\ln a$. The effective equation of state parameter can also be defined as
\begin{align}
\omega_{eff}\equiv-\f13-\f12\f{h^\prime}{h^2}=-1+\Omega_m+\f43\Omega_r.
\end{align}
The fixed points of the above system is collected in table \ref{tab1}.
\begin{table}[h!]
	\begin{center}
		\begin{tabular}{|c||c|c|c|c|}
			\hline
			~&\textbf{fixed point}&\textbf{$\omega_{eff}$}&\textbf{condition}&\textbf{stability}\\
			\hhline{|=||=|=|=|=|}
			$P_r$&radiation&1/3&$\eta=0$ and $\eta\geq1$&saddle\\
			\hline
			$P_m$&dust&0&$\eta=0$ and~$\eta>1$&saddle\\
			\hline
			$P_\Lambda$&de Sitter&-1&all $\eta$&stable\\
			\hhline{|=||=|=|=|=|}
			~&de Sitter&-1&$\eta=1,\xi=1/4$&stable\\
			\hhline{|~||=|=|=|=|}
			$P_1$&stiff matter&1&$\eta=1,\xi=1/2$&saddle\\
			\hhline{|~||=|=|=|=|}
			~&dust&0&$\eta=1,\xi\rightarrow0$&saddle\\
			\hhline{|~||=|=|=|=|}
			~&radiation&1/3&$\eta=1,\xi\rightarrow\infty$&saddle\\
			\hline
		\end{tabular}
	\end{center}
	\caption{Fixed points of the dynamical system \eqref{dy1}-\eqref{dy3}.\label{tab1}}
\end{table}
\subsection{Radiation fixed point: $P_r$}
The first fixed point of the dynamical system \eqref{dy1}-\eqref{dy3} is
\begin{align}
P_r=(\Omega_m,\Omega_r,\Omega_h)=\left(0,1,0\right).
\end{align}
The effective equation of state parameter can be calculated as $\omega_{eff}=1/3$, showing that this fixed point is a radiation dominated point. The eigenvalues associated with this fixed point is
\begin{align}
\left(-4,0,5\right),
\end{align}
indicating that the radiation fixed point is saddle. This fixed point exists for $\eta\geq1$ and $\eta=0$.
\subsection{Dust fixed point: $P_m$}
For this fixed point we have
\begin{align}
P_m=(\Omega_m,\Omega_r,\Omega_h)=\left(1,0,0\right),
\end{align}
and we have $\omega_{eff}=0$ showing that $P_m$ describe a dust dominated universe. The eigenvalues of this fixed point is
\begin{align}
(-3,3,-1),
\end{align}
implying that the dust dominated fixed point is also a saddle point. This fixed point exists only for $\eta>1$ and also $\eta=0$.
\subsection{de Sitter fixed point: $P_\Lambda$}
This fixed point corresponds to
\begin{align}
P_\Lambda=(\Omega_m,\Omega_r,\Omega_h)=\left(0,0,y\right),
\end{align}
where $y$ is an arbitrary constant for $\eta\geq1$ and an arbitrary non-vanishing constant for $\eta<1$. For this fixed point we have $\omega_{eff}=-1$ implying that this fixed point describes a de Sitter expanding phase. The eigenvalues of this fixed point is
\begin{align}
\left(0,-4,-4+\f{1}{1-3\xi\eta(12-y)^{\eta-1}}\right).
\end{align}
In order to have a stable de Sitter fixed point, the constant $y$ should satisfy the relation
\begin{align}
(12-y)^{\eta-1}<(4\xi\eta)^{-1}.
\end{align}
\subsection{The fixed point: $P_1$}
This fixed point only exists for $\eta=1$ which is the original Rastall theory. This fixed point corresponds to
\begin{align}
P_1=(\Omega_m,\Omega_r,\Omega_h)=\left(\f{1-4\xi}{1-3\xi},0,0\right),
\end{align}
with the equation of state parameter $\omega_{eff}=\xi/(3\xi-1)$. The eigenvalues of this fixed point is
\begin{align}
\left(\f{1}{3\xi-1},\f{5-12\xi}{3\xi-1},\f{12\xi-3}{3\xi-1}\right),
\end{align}
In the case $\xi=1/4$ we have $\omega_{eff}=-1$ corresponding to the de Sitter expanding universe. In this case the eigenvalues are $(-4,0,0)$, and we can show that the de Sitter fixed point is stable. In the case $\xi=1/2$ we have $\omega_{eff}=1$ with eigenvalues $(2,6,-6)$ which corresponds to a saddle stiff matter phase. For very large values of $\xi$ we have $\omega_{eff}\rightarrow1/3$ with eigenvalues $(0,-4,4)$ which is a saddle radiation dominated fixed point. At last, for $\xi\rightarrow0$ we have $\omega_{eff}\rightarrow0$ with eigenvalues $(-1,-3,3)$ indicating a saddle dust dominated fixed point.

In summary, one can see from the above calculations that for $\eta<1$ there exists only one fixed point, and behaves as a stable de Sitter expanding universe. As a result the $\eta<1$ case could not fully address the history of the universe. 

In the case of Rastall theory $\eta=1$ there are three fixed point, one of them is de Sitter expanding and the other is a radiation dominated fixed point. The third one can be de Sitter, dust or radiation dominated depending on a specific value of the parameter $\xi$. The original Rastall theory could then explain the history of the universe. The universe could start from the radiation dominated fixed point which will go to the dust dominated point which corresponds to very small values of $\xi$. At the end the universe will stay at the stable de Sitter fixed point $P_\Lambda$. 

In the case $\eta>1$, there are three fixed points corresponding to a saddle dust, saddle radiation and stable de Sitter. In this case we do not have any constraint on the value of $\xi$. As a result in this case, the history of the universe could be obtained consistently for all values of the parameter $\xi$. In this sense, the generalized Rastall theory with $\eta>1$ would be more preferable.

 It should be noted that for $\eta>1$ and $\eta=0$, the universe has its standard radiation, matter and de Sitter phase. So, qualitatively astrophysical considerations should be satisfactory in this model. However, early time behaviors as well as structure formation should be analyzed separately to make more precise conclusions.
\section{Matter perturbations}\label{cosmopert}
\begin{table*}
	\begin{center}
		\begin{tabular}{|c||c|c||c|c||c|c||c|c||c|}
			\hline
			$~~~\eta~~~$&$~~~~\xi~~~~$&$1\sigma/2\sigma$~intervals&$~~~~\sigma_8^0~~~~$&$1\sigma/2\sigma$~intervals&$~~~~h_0~~~~$&$1\sigma/2\sigma$~intervals&$~~~~\zeta~~~~$&$1\sigma/2\sigma$~intervals&$\chi^2/dof$\\
			\hhline{|=||=|=||=|=||=|=||=|=||=|}
			\multirow{2}{*}{0.6}&\multirow{2}{*}{0.435}&$\pm0.156$&\multirow{2}{*}{0.800}&$\pm0.019$&\multirow{2}{*}{0.968}&$\pm0.020$&\multirow{2}{*}{0.920}&$\pm0.009$&\multirow{2}{*}{0.44}\\
			\hhline{|~||~|-||~|-||~|-||~|-||~|}
			~&~&$\pm0.307$&~&$\pm0.038$&~&$\pm0.039$&~&$\pm0.018$&~\\
			\hhline{|=||=|=||=|=||=|=||=|=||=|}
			\multirow{2}{*}{1}&\multirow{2}{*}{0.269}&$\pm0.0.082$&\multirow{2}{*}{0.719}&$\pm0.017$&\multirow{2}{*}{0.977}&$\pm0.020$&\multirow{2}{*}{1.006}&$\pm0.010$&\multirow{2}{*}{0.47}\\
			\hhline{|~||~|-||~|-||~|-||~|-||~|}
			~&~&$\pm0.161$&~&$\pm0.034$&~&$\pm0.040$&~&$\pm0.019$&~\\
			\hhline{|=||=|=||=|=||=|=||=|=||=|}
			\multirow{2}{*}{1.3}&\multirow{2}{*}{0.263}&$\pm0.032$&\multirow{2}{*}{0.669}&$\pm0.016$&\multirow{2}{*}{1.003}&$\pm0.021$&\multirow{2}{*}{1.092}&$\pm0.013$&\multirow{2}{*}{0.45}\\
			\hhline{|~||~|-||~|-||~|-||~|-||~|}
			~&~&$\pm0.063$&~&$\pm0.031$&~&$\pm0.041$&~&$\pm0.025$&~\\
			\hline
			\multirow{2}{*}{$\eta=0~(\Lambda CDM)$}&\multirow{2}{*}{-}&\multirow{2}{*}{-} &\multirow{2}{*}{0.81}&$\pm0.0061$&\multirow{2}{*}{1.04}&$\pm1.7$&\multirow{2}{*}{-}&\multirow{2}{*}{-}&\multirow{2}{*}{0.7}\\
			\hhline{|~||~|~||~|-||~|-||~|~||~|}
				~&~&~&~&$\pm0.012$&~&$\pm2.8$&~&~&~\\
				\hline
		\end{tabular}
	\end{center}
	\caption{Best fit values of the model parameters $h_0=H_0/70$, $\xi$, $\zeta$ and $\sigma_8^0$ together with their $1\sigma$ and $2\sigma$ confidence intervals for three different values of $\eta=0.6,1,1.3$.\label{tab2}}
\end{table*}
In this section, we will consider the evolution of the matter density perturbations for Rastall-$f(R)$ theory. The scalar perturbation of the metric in Newtonian gauge ($B=0=E$) can be written as
\begin{align}
ds^2=a^2\Big[-(1+2\varphi)dt^2+(1-2\psi)d\vec{x}^2\Big],
\end{align}
where $\varphi$ and $\psi$ are the Bardeen potentials.
The scalar perturbation of the energy momentum tensor can be written as
\begin{align}
&\delta T^0_0=-\delta\rho\equiv-\rho\,\delta,\nonumber\\& \delta T^0_i=(1+w)\rho\,\partial_i v,\quad \delta T^i_j=\delta^i_j\delta p.
\end{align}
Here, $\delta$ is the matter density contrast defined as $\delta=\delta\rho/\rho$, $v$ is the scalar mode of the velocity perturbation of the fluid and $\delta p$ is the pressure perturbation. Also $w$ is the unperturbed equation of state parameter, $w=p/\rho$.
In this paper we will assume that the unperturbed matter content of the Universe are of dust form $w=0$.

The first order perturbation of $(i\neq j)$ components of the Einstein field equation \eqref{eqeins} can be obtained as
\begin{align}
\varphi=\psi.
\end{align}
This implies that in the generalized Rastall theory the anisotropy factor $\eta_a\equiv\varphi/\psi$ is equal to unity which is similar to the standard Einstein theory. The $(00)$, $(0i)$ and $(ii)$ components of the Einstein field equation can be written as
\begin{align}\label{00comp}
a^2\rho\delta_m&+4\kappa^2(k^2+3H^2)\varphi+4\kappa^2\Big[3H\dot\varphi\nonumber\\&-(k^2\varphi+6H^2\varphi+6\dot H\varphi+12H\dot\varphi+3\ddot\varphi)f^\prime\Big]=0
\end{align}
\begin{align}\label{zeroi}
\rho\theta-4\kappa^2\f{k^2}{a^2}(\dot{\varphi}+H\varphi)=0,
\end{align}
and
\begin{align}\label{ii}
\ddot\varphi&+3H\dot\varphi+(2\dot{H}+H^2)\varphi\nonumber\\&-3f^\prime\left[\ddot\varphi+4H\varphi+2(H^2+\dot{H})\varphi+\f13k^2\varphi\right]=\f{a^2\delta p}{4\kappa^2}.
\end{align}
In obtaining the above perturbed equations, we have Fourier transformed the perturbed fields. Also we have defined the Fourier transformed velocity divergence as $\theta=-k^2v$. 

The Fourier transformed first order perturbations of the conservation equation \eqref{eqcons}, can be simplified to
\begin{widetext}
\begin{align}\label{47}
a^4\rho\dot\delta+3a^4 H\delta p&+a^4\Big[\dot\rho\delta+\rho(3H\delta+\theta-3\dot\varphi)\Big]+4\kappa^2a^2\Big[12H^3\varphi-(k^2-18(H^2+\dot{H}))\dot\varphi-6\ddot{H}\varphi-6H\ddot\varphi-\dddot\varphi\Big]f^\prime\nonumber\\&+24\kappa^2(2H^3-\ddot{H})\Big[(k^2+6(H^2+\dot{H}))\varphi+12H\dot\varphi+3\ddot\varphi\Big]f^{\prime\prime}=0,
\end{align}
\begin{align}\label{48}
\dot\rho\theta+\rho(4H\theta-k^2\phi+\dot\theta)-k^2\delta p-4\kappa^2\f{k^2}{a^2}\Big[(k^2+6(\dot{H}+H^2))\varphi+12H\dot\varphi+3\ddot\varphi\Big]f^\prime=0.
\end{align}
\end{widetext}
Now, let us consider the sub-horizon limit of the theory where $k\gg aH$, where $k$ is the wave number. The (00) component of the metric field equation \eqref{00comp} becomes
\begin{align}\label{ppp}
a^2\rho\delta+4\kappa^2 k^2(1-f^\prime)\varphi=0.
\end{align}
It should be noted that in the subhorizon limit the expansion of the universe can be neglected. As a result only the term proportional to $\delta$, $\theta$, $\delta p$ and $k^2$ survive in the field equation.
From the $(ii)$ component of the metric field equation \eqref{ii} at subhorizon limit one obtains
\begin{align}\label{pp}
\delta p=-\frac{4\kappa^2}{a^2}k^2f^\prime\varphi.
\end{align}
The $(0i)$ component of the metric equation \eqref{zeroi} is not independent and can be obtained from \eqref{ppp} and \eqref{48}.

By solving equation \eqref{47} for $\theta$, substituting the result in \eqref{48}, using \eqref{pp} and performing the subhorizon limit, one can obtain the dynamical evolution equation of the matter density contrast in the subhorizon limit as
\begin{widetext}
\begin{align}
\ddot\delta&+\left(7H+2\frac{\dot{\rho}}{\rho}\right)\dot\delta+\left(12H^2+3\dot H+7H\f{\dot\rho}{\rho}+\f{\ddot\rho}{\rho}\right)\delta+\f{8\kappa^2}{a^2\rho}k^2\Bigg[(k^2+3H^2+3\dot H)\varphi+5H\dot\varphi+\ddot\varphi\Bigg]f^\prime\nonumber\\&-\f{24\kappa^2}{a^4\rho}k^2\Bigg[2(\ddot H-2H^3)\dot\varphi+2(\dddot H-6H^2\dot H)\varphi\Bigg]f^{\prime\prime}+k^2\varphi\Big[1-\f{144\kappa^2}{a^6\rho}(2H^3-\ddot H)^2f^{\prime\prime\prime}\Big]=0.
\end{align}
\end{widetext}
Now, substituting $\varphi$ from equation \eqref{ppp} and using background field equations \eqref{consfrw} and \eqref{frid} to substitute $\rho$ and derivatives of $H$ in terms of the Ricci scalar $R$, one can obtain the dynamical evolution equation for matter density perturbation as
\begin{align}\label{eq3}
\ddot\delta+B_\delta\dot\delta+C_\delta=0,
\end{align}
where we have defined
\begin{align}
B_\delta=\f{(2-F^\prime)^2+6F^{\prime\prime}(R+4F)}{3F^\prime(F^\prime-2)}H,
\end{align}
\begin{align}
\f{C_\delta}{a^2}&=\f{F(4F+R)^2}{3(F^\prime-2)F^{\prime2}}F^{\prime\prime\prime}+\f{F(4F+R)^2(2-3F^\prime)}{3(2-F^\prime)^2F^{\prime3}}F^{\prime\prime2}\nonumber\\&-\f{(4F+R)(4F(F^\prime-1)+(10F+R)F^{\prime2})}{6(2-F^\prime)F^{\prime3}}F^{\prime\prime}\nonumber\\&+\f{1}{18}\Bigg[14F+R+12\left(1+\f{1}{F^\prime}\right)\f{k^2}{a^2}\nonumber\\&\qquad\qquad~~\,+\f{2}{F^{\prime2}}\Big((1+2F^\prime)F-RF^\prime\Big)\Bigg],
\end{align}
where	
\begin{align}
F=3(\Lambda+f)-R.
\end{align}
As before, in the following, we will consider the special case $f=\epsilon R^\eta$ where $\epsilon$ and $\eta$ are constants. Also, from now on, we will transform the above equations to redshift coordinates and analyze the dynamics of the system as a function redshift $z$.

In order to compare the model with the observational data, we use the data on $f\sigma_8$ \cite{obs} defined as
\begin{align}
f\sigma_8(z)=\sigma_8(z)\f{d\ln \delta(z)}{d\ln a},\qquad \sigma_8(z)=\sigma_8^0\f{\delta(z)}{\delta(0)},
\end{align} and also the $H_0$ data set \cite{obshubble} in the redshift range $0<z<2$.
\begin{figure}[h!]
	\includegraphics[scale=0.545]{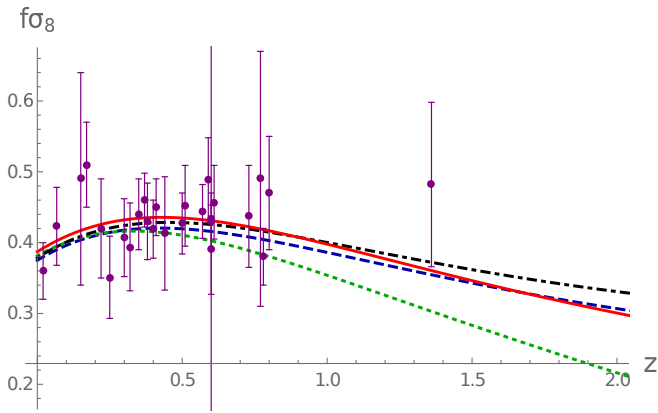}
	\caption{\label{fsigma8}The evolution of the quantity $f\sigma_8$ as a function of redshift for different values of $\eta=0.6$ (dotted), $1$(dashed), $1.3$ (dot-dashed). The solid red line corresponds to the conservative $\Lambda$CDM theory. The error bars indicate the observational values \cite{obs}. In plotting the figures, we have assumed that $\gamma\equiv k/H_0=1.4$.}
\end{figure}
We will numerically solve the set of equations \eqref{eq1}, \eqref{eq2} and \eqref{eq3} with the initial conditions $h(0)=1$, $\Omega_m(0)=\Omega_{m0}=0.305$, $\delta(0)=1$ and $\delta^\prime(0)=\zeta \delta_{0}$, where $\delta_0=-0.517$ is the $\Lambda$CDM value of the current derivative of the matter density contrast, and $\zeta$ is a constant which indicates deviations from the $\Lambda$CDM value. In the following, we will estimate the best fit values of the parameters $\xi$, $H_0$, $\sigma_8^0$ and $\zeta$.
In order to do this, we use the likelihood analysis of the model based on the data on $H_0$ and $f\sigma_8$.

 In the case of independent data points, the likelihood function can be defined as
\begin{align}
L=L_0e^{-\chi^2/2},
\end{align}
where $L_0$ is the normalization constant and the quantity $\chi^2$ is defined as
\begin{align}
\chi^2=\sum_i\left(\f{O_i-T_i}{\sigma_i}\right)^2.
\end{align}
Here $i$ counts the data points, $O_i$ are the observational value, $T_i$ are the theoretical values and $\sigma_i$ are the error associated with the $i$th data from observation. 
For the generalized Rastall theory, the likelihood function can be defined as
\begin{align}
L=L_0\,\textmd{exp}\left[-\f12\sum_i\left(\f{O_i-\sigma_8^0T_i}{\sigma_i}\right)^2\right],
\end{align}
By maximizing the likelihood function one can find the best fit values of the parameters. In table \ref{tab2}, we have summarized the result of the maximum likelihood estimation on the parameters $\xi$, $h_0=H_0/70$, $\sigma_8^0$ and $\zeta$ for different values of $\eta=0.6,\,1,\,1.3$ together with their $1\sigma$ and $2\sigma$ confidence interval. It should be noted that the value of $\xi$ is positive up to $2\sigma$ level for all values of $\eta$. In figure \eqref{fsigma8}, we have plotted the evolution of $f\sigma_8$ as a function of the redshift. The red solid line denotes the $\Lambda$CDM curve. One can see that the generalized Rastall theory could explain the observational data especially for small values of the redshift where we have more data. For larger values of the redshift, the evolution of the $f\sigma_8$ function differs from its $\Lambda$CDM counterpart. The smaller values of the $\eta$ parameter, leads to the smaller values of $f\sigma_8$ at larger redshifts. This suggests that more data on the $f\sigma_8$ quantity on redshifts $z>1$ would be needed to fully decide which theory could satisfy the observations better. In table \ref{tab2}, we have also reported $\chi^2/dof$ of the $\sigma_8^0$ estimation. One can also deduce from the value of this quantity that more observational data would be needed to make a good decision about the generalized Rastall theory.
 \section{conclusion}
In this paper, we have considered the cosmological implications of the generalized Rastall theory, where the Ricci scalar in the (non-)conservation equation is substituted by an arbitrary function of the Ricci scalar $f(R)$. In order to make quantitative arguments, we have assumed that the function $f$ has a power-law form as $f=\epsilon R^\eta$. The cosmological equations has been obtained and we have shown that the generalized Rastall theory could explain the observational data on the Hubble parameter. The larger values of $\eta$ imply smaller values of the Hubble parameter $H$ for larger reshifts. It should be noted that the $\eta=0$ case is identical to the standard $\Lambda$CDM model. As a result one can obtain a series of models with varying $\eta\geq0$; The upper curve in the Hubble plot corresponds to the smallest value of $\eta$. The evolution of the deceleration parameter also shows that the universe described by the generalized Rastall theory would predict more acceleration for the redshift range $z>0.2$ compare to the $\Lambda$CDM theory. Larger values for $\eta$ would imply more acceleration of the universe.

We have also considered the dynamical system analysis of the theory. In this way, we have found that there are two classes of fixed points, one for the original Rastall theory with $\eta=1$ and the other class is for arbitrary values of $\eta$. For general $\eta$, we have always a Stable de Sitter and unstable radiation fixed points. However, the dust dominated fixed point does not exist for $0<\eta<1$. It should be note that the case $\eta=0$ which is identical to Einstein's gravity, we have three standard dust, radiation and de Sitter fixed points. For the case of Rastall theory $\eta=1$, beyond stable de Sitter and unstable radiation fixed points, there is another fixed point with equation of state parameter $\omega_{eff}=\xi/(3\xi-1)$. Depending on the value of $\xi$, this fixed point can behave as dust, radiation, de Sitter and also stiff matter node. In summary, for $\eta\geq1$ and $\eta=0$, the history of the universe can be explained satisfactorily in this model. However, for $0<\eta<1$, the dust dominated fixed point is absent from the system and the universe will not stay enough time in that stage. As a result, the case $0<\eta<1$ should be eliminated from the parameter space of the theory.

In this paper, we have also considered the growth of matter perturbations in the generalized Rastall theory. We have found that the anisotropic stress in this theory is equal to unity which is the same as Einstein general relativity. Also, due to the non-conservative nature of the theory, the evolution equation of the matter density contrast is modified in this model. In order to compare the theory with observational data, we have considered the evolution of $f\sigma_8$ and find the best fit values of the model parameters $\xi$ and $\zeta$ and cosmological parameters $H_0$ and $\sigma_8^0$ using the maximum likelihood analysis on the $H_0$ and $f\sigma_8$ data sets. One can see from the evolution plot of $f\sigma_8$ that for redshifts smaller that unity where we have more data, the generalized Rastall theory predict the same evolution of $f\sigma_8$ as $\Lambda$CDM theory. For larger redshifts, the evolution of $f\sigma_8$ will differ from the $\Lambda$CDM value. The value of $\chi^2/dof$ for various $\eta$'s suggests that more data would be needed in order to decide which theory satisfy the observations better. Finally it should be noted that in this paper, we have used local observational data \cite{obs,obshubble}. As one can see from table II, the $\sigma_8^0$ value is roughly $0.7$, so the model could not make the $\sigma_8$ tension better. Also, from the table, one can see that the Hubble parameter is around $70$. This shows that model can make the $H_0$ tension better. However, for precise statement about this issue, one needs a full set of observational data which is the scope of the future works.

\end{document}